\documentclass[doublecol]{epl2}

\usepackage{amssymb,amsfonts,bm}
\usepackage{graphics,graphicx}

\title{Ground state of classical bilayer Wigner crystals}

\author{Ladislav \v{S}amaj\inst{1,2} and Emmanuel Trizac\inst{2}}
\institute{
\inst{1}Institute of Physics, Slovak Academy of Sciences, Bratislava, Slovakia\\
\inst{2}Universit\'e Paris-Sud, Laboratoire de Physique Th\'eorique et 
Mod\`eles Statistiques, UMR CNRS 8626, 91405 Orsay, France.
}
\abstract{We study the ground state structure of electronic-like bilayers, 
where different phases compete upon changing the inter-layer separation 
or particle density.
New series representations with exceptional convergence properties are derived
for the exact Coulombic energies under scrutiny. 
The complete phase transition scenario --including critical phenomena-- can 
subsequently be worked out in detail, thereby unifying a rather scattered 
or contradictory body of literature, hitherto plagued by 
the inaccuracies inherent to long range interaction potentials.}


\begin{document}

\maketitle

The prediction by Wigner that strongly correlated charge carriers
in a uniform compensating background could crystallize \cite{Wi34},
was first realized experimentally with electrons at the surface 
of liquid Helium \cite{GrAd79}, which form a two-dimensional structure.
Since then, the study of low dimensional electronic systems has shown 
no abating and in particular, the bilayer geometry singles out.
It appears significantly richer than its monolayer counterpart and
has been investigated in different settings~:
GaAs quantum wells \cite{Shaye} or other semiconductors \cite{Eise04},
quantum dots \cite{ImMA96}, graphene \cite{graphene}, boron nitride
\cite{boron}, laser-cooled trapped ion plasmas \cite{Mitc98},
dusty plasmas \cite{dusty} and colloids \cite{colloids,rque0}.
In light of these applications, it is essential to understand the
ground state features of Coulombic bilayers, starting with the 
classical limit. This problem has received significant attention in the last
20 years, as such \cite{Falk94,EsKa95,GoPe96,Messina03,LoNe07} 
or supplemented with finite temperature
analysis \cite{ScSP99,WeLJ01,Maza11}. 
In addition, ground state ordering impinges on 
strong-coupling expansions describing counterintuitive
yet ubiquitous electrostatic phenomena, such as like-charge
attraction \cite{Levin02,Naji,LaLP00,SaTr11} 
or charge reversal \cite{Grosberg02,TrVa06}.
This body of work has revealed the main features of ground state structure,
but there exist surprising discrepancies in the literature, 
especially on the respective domains of existence of the different phases 
possible.
%as discussed below.
The reason lies in the long range nature of Coulombic interactions,
a common bane for such analysis. Our goal here is to resolve existing
controversies, precisely locate all phases and discuss the critical
behavior associated. 
%to ordering for a Wigner bilayer. 
All results reported
are exact; they are obtained from new series representations of lattice
sums for Coulomb law. 

\begin{figure}[tb]
\begin{center}
\includegraphics[width=0.4\textwidth]{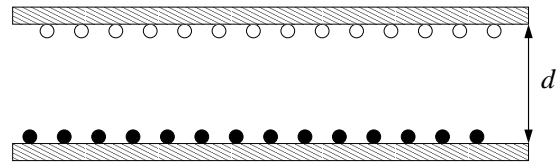}
\caption{{Side view of the two parallel plates at a separation $d$. 
From their common uniform
surface charge density $-\sigma q$, we define the dimensionless distance
as $\eta = d \sqrt{\sigma}$. Ions, which  bear charge $q$, are shown 
as black or white disks for visual ease, but they are point-like in the present 
study.} }
\label{fig:geom}
\end{center}
\end{figure}

\begin{figure}[tb]
\begin{center}
\includegraphics[width=0.4\textwidth]{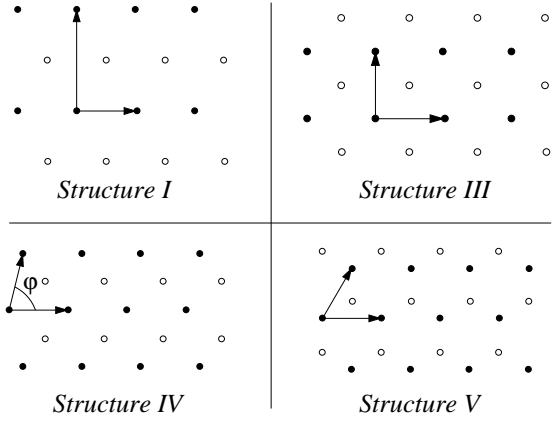}
\caption{Schematic representation of the different ground states 
encountered when the dimensionless distance $\eta$ increases. The open and 
filled symbols show the locations of ions on the opposite surfaces
(see Fig.  \ref{fig:geom}). The arrows
are for lattice vectors $\bm{a}_1$ and $\bm{a}_2$, from which we define
the aspect ratio $\Delta=|\bm{a}_2|/|\bm{a}_1|$: $\Delta=\sqrt 3$ and $1$
for structures I and III respectively, while structure II interpolates 
between I and III with $1<\Delta<\sqrt 3$. For structure IV, 
the order parameter is the angle $\varphi$ between $\bm{a}_1$ and $\bm{a}_2$. 
We have $\varphi=\pi/2$ for structure III whereas
$\varphi=\pi/3$ for structure V, and in general,
$\pi/3 \leq \varphi \leq \pi/2$. Structures I, III and V are rigid,
as opposed to the soft cases II and IV where the unit cell geometry depends
on inter-plate separation, through $\Delta$ and $\varphi$, respectively.
Note that the shift between the two
opposite crystals materialized by open and filled symbols is 
$ (\bm{a}_1+\bm{a}_2)/2$ for structures I, II, III, IV but 
$(\bm{a}_1+\bm{a}_2)/3$ for structure V.}
\label{fig:struct}
\end{center}
\end{figure}

We consider an ensemble of identical classical point charges $q$, interacting 
through a $1/r$ pair potential, and confined between two symmetric parallel
charged walls. These boundaries both bear a uniform surface charge 
of density $-\sigma q$, so that global electroneutrality holds. 
At finite temperature $T$, the charges do populate the
interior of the slab. For $T=0$ though, the charges 
evenly condense on the opposing walls, thereby
forming a bilayer ground state \cite{rque,rque101} 
the structure of which depends on a single dimensionless parameter
$\eta = d \sqrt{\sigma}$, where $d$ is the inter-plate distance
{(see Fig. \ref{fig:geom})}. 
It is known that five structures
can be realized upon increasing $\eta$; they will be referred to following
standard terminology, common to the classical \cite{GoPe96} and 
quantum contexts \cite{NaHo95}. To begin with, the limits
of small and large $\eta$ are both straightforward. For
$\eta \to 0$, a genuine two-dimensional one component plasma is produced
\cite{Levin02}, where the strong mutual repulsion between charges
leads to a triangular Wigner crystal \cite{BoMa77}, the so-called structure I. 
Conversely, for $\eta \to \infty$, the two layers decouple and 
a hexagonal crystal forms on each plate (structure V).
%, with particle density $\sigma$ as opposed to $2\sigma$ for structure I. 
These two crystals adopt
a staggered configuration, to minimize inter-layer repulsion. For 
intermediate reduced distances $\eta$, three other structures are met, see
Fig. \ref{fig:struct}: a staggered rectangular lattice (structure II),
a staggered square lattice (structure III), and a staggered rhombic lattice
(structure IV).
Note that while one can evolve continuously through the sequence 
I $\to$ II $\to$ III $\to$ IV, no continuous deformation allows
to create structure V from one of the others. 
The transitions between phases will therefore be of different orders,
with characteristics and critical exponent (in the continuous cases)
that will be worked out explicitly below. A goal of our 
analysis is to precisely locate the transition points between phases:
indeed, a dispersion of about 20\% exists for the hitherto reported threshold
$\eta_{\mathrm{\tiny IV}}$ between structures IV and V, see 
\cite{EsKa95,GoPe96,WeLJ01,LoNe07}.
In addition, controversial results have been reported for the transition 
point $\eta_{\mathrm{\tiny I}}$ between structures I and II: 
$\eta_{\mathrm{\tiny I}} \simeq 0.006$ from Ewald summation technique
\cite{GoPe96}, $\eta_{\mathrm{\tiny I}} \simeq 0.028$ from Monte Carlo
simulations \cite{WeLJ01}, whereas lattice sum minimization
of Yukawa systems in the unscreened limit hints at $\eta_{\mathrm{\tiny I}} = 0$
\cite{Messina03}, meaning that structure I could possibly only exist
at precisely $\eta=0$. 

We start by addressing the question whether $\eta_{\mathrm{\tiny I}} = 0$ or 
$\neq0$, and to this end, we compute the energy $E(\Delta,\eta)$ of 
structure II. 
For a given layer, the 2D lattice points are indexed by 
$j \bm{a}_1 + k \bm{a}_2$ where $j$ and $k$ are integers and 
the lattice vectors $\bm{a}_1 = a (1,0)$, $\bm{a}_2 = a (0,\Delta)$ 
are shown in bold in Fig. \ref{fig:struct}. 
The global electroneutrality requires that $a^2\sigma\Delta=1$. 
The aspect ratio $\Delta$ fulfills $1\leq \Delta \leq \sqrt{3}$ 
with $\Delta=\sqrt{3}$ for structure I and $\Delta =1$ for the (square) 
structure III. 
The dielectric constant of the medium is set to unity for the sake 
of simplicity, and the total energy per ion $E(\Delta,\eta)$ is written 
as the sum of intra- and inter-layers contributions.
%We provide here the key steps leading to the intra-layer term, from
We first restrict ourselves to a disk of finite radius $R$ around 
a given reference ion located at $(0,0)$. 
Considering ion-ion and ion-plate interactions, we have the intra-layer energy
\begin{equation}
E_{\rm intra}=\frac{q^2}{2a} \sum_{j,k\atop (j,k)\ne (0,0)}
\frac{1}{\sqrt{j^2 + k^2 \Delta^2}} 
-\frac{\sigma q^2}{2} \int_0^R d{\bf r} \frac{1}{\vert{\bf r}\vert}
\end{equation}
with the restriction $j^2 +k^2 \Delta^2 \le \left( R/a \right)^2 $. 
It is expedient and common procedure \cite{dLPS80,Maza11}
to use the gamma identity 
$(\pi/z)^{1/2} = \int_0^\infty t^{-1/2} \exp(-zt)\, dt$ valid for $z>0$,
which provides us with a simple expression where the limit $R\to\infty$ can be 
readily taken:
\begin{eqnarray}
\frac{\sqrt{\pi}E_{\rm intra}}{q^2\sqrt\sigma} 
= \frac{1}{2a\sqrt\sigma} \int_0^{\infty} \frac{dt}{\sqrt{t}} 
\left[ \sum_{j,k} e^{-t j^2} e^{-t k^2\Delta^2} 
- 1 - \frac{\pi}{t\Delta}  \right] \nonumber\\
= \frac{1}{2}\int_0^{\infty} \frac{dt}{\sqrt{t}} 
\left[ \sum_{j,k} e^{-t j^2/\Delta} e^{-t k^2\Delta} 
- 1 - \frac{\pi}{t}  \right], \nonumber\\
= \int_0^{\pi} dt \frac{\pi}{t^{3/2}} \left[
\sum_{j,k \neq (0,0)} e^{-(\pi j)^2/(\Delta t)} e^{-(\pi k)^2\Delta/t}  \right] - 
2\sqrt{\pi} . \nonumber \\
\end{eqnarray}
Here, the second line is obtained from the substitution 
$t\Delta \to t$ and the condition $a^2 \sigma \Delta=1$;
the third line stems from considering separately the domains
$t\in[0,\pi]$ and $t\in [\pi,\infty]$ in the integral, 
substituting $\pi^2/t \to t$ and subsequently using 
Poisson summation formula
\begin{equation} 
\sum_{j=-\infty}^{\infty} e^{-(j+\phi)^2 t} = \sqrt{\frac{\pi}{t}}
\sum_{j=-\infty}^{\infty} e^{2\pi ij\phi}e^{-(\pi j)^2/t} .
\label{eq:Poisson}
\end{equation} 
The inter-layer energy contribution $E_{\rm inter}$ 
is amenable to a similar treatment
\cite{SaTr12}, and the last step of the procedure consists in introducing
the function 
\begin{equation} 
z_{\nu}(x,y) = \int_0^{1/\pi} \frac{dt}{t^{\nu}} e^{-x t} e^{-y/t} \qquad
\mbox{for } y>0.
\label{eq:z}
\end{equation}
We finally end up with the series representation for the total energy
$E(\Delta,\eta)= E_{\rm intra} + E_{\rm inter} $:
\begin{widetext}
\begin{eqnarray}
\frac{E(\Delta,\eta) \,\sqrt{\pi}}{q^2\sqrt{\sigma}} & = & 
 2\sum_{j=1}^{\infty} 
\left[ z_{3/2}(0,j^2/\Delta) + z_{3/2}(0,j^2\Delta) \right]
+ 4 \sum_{j,k=1}^{\infty} z_{3/2}(0,j^2/\Delta+k^2\Delta) \nonumber \\ & &
+  \sum_{j=1}^{\infty} (-1)^j 
\left[ z_{3/2}((\pi\eta)^2,j^2/\Delta) + z_{3/2}((\pi\eta)^2,j^2\Delta) \right]
+ 2 \sum_{j,k=1}^{\infty} (-1)^j (-1)^k z_{3/2}((\pi\eta)^2,j^2/\Delta+k^2\Delta) 
\nonumber \\ & &
+ 2 \sum_{j,k=1}^{\infty} z_{3/2}(0,\eta^2+(j-1/2)^2/\Delta+(k-1/2)^2\Delta)
- 2 \sqrt{\pi} - \frac{\pi}{2} z_{1/2}(0,\eta^2)  . 
\label{eq:series1}
\end{eqnarray}
\end{widetext}
The function $z_\nu$ generalizes to two-layer problems the so-called
Misra function \cite{Misra40} used extensively 
in single-layer lattice summation \cite{Borwein88,Bowick06}.
Our use of (\ref{eq:series1}) will be three-pronged: 
it allows to show analytically that $\eta_{\mathrm{\tiny I}} = 0$,
to calculate explicitly the singular behavior near critical points
and is moreover particularly suited for numerical evaluations.
From an operational point of view, the series (\ref{eq:series1}) is indeed 
endowed with remarkable properties: it is free of singular terms, and 
importantly, converges extremely quickly. The error made upon truncating
the series in the energy expression (\ref{eq:series1}) at order 
$j=k=M$ behaves like $\exp(-c M^2)/M$, where $c$ is a constant of order unity.
We first document the convergence property on the single-layer case
of structure I, for which the exact energy is 
$E(\sqrt 3,0)/(q^2\sqrt{2\sigma}) = -1.96051578931989165\ldots$,
which is directly the Madelung constant of the 2D hexagonal Wigner
crystal. 
Cutting the series (\ref{eq:series1}) at $M$, we obtain 
the exact value with a precision of $2,5,10,17$ digits with 
notably small cutoffs $M=1,2,3,4$ respectively. This makes numerical
calculations extremely fast and efficient on any workstation.
A similar accuracy is met for all other structures and parameter values 
reported here,
and all numerical results quoted below have been obtained with 
the cutoff $M=5$.

\begin{figure}[htb]
\includegraphics[width=0.4\textwidth]{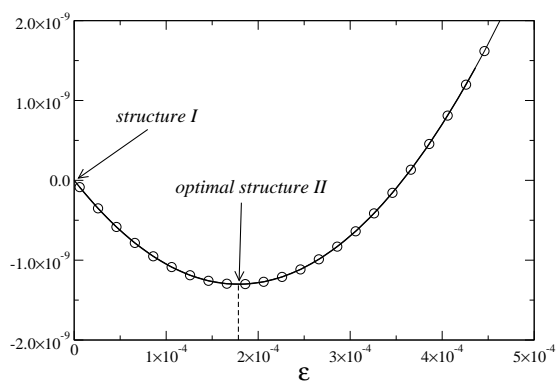} 
\caption{The difference between the dimensionless energies 
$[E(\Delta,\eta)-E(\sqrt{3},\eta)]/(q^2\sqrt{2\sigma})$ versus 
$\epsilon=\sqrt{3}-\Delta$, for $\eta=5.\,10^{-3}$. The analytical formula
(\ref{eq:epsilon1}) with the Taylor expansions of $f_1$ and $f_2$ given 
in the text is shown by the continuous line. It is compared to
the numerical evaluation of the series
(\ref{eq:series1}) with a cutoff $M=5$ (symbols). The optimal $\epsilon$ 
value following from the prediction (\ref{eq:transitionItoII})
is shown by the dashed vertical line.}
\label{fig:enI}
\end{figure}

We now turn our attention to the threshold $\eta_{\mathrm{\tiny I}}$
which defines the stability window of structure I. For a given distance
$\eta$, we proceed by calculating the Taylor expansion of (\ref{eq:series1})
in the small parameter $\epsilon = \sqrt3 -\Delta$, which yields
\begin{equation}
\frac{E(\sqrt{3}-\epsilon,\eta)}{q^2\sqrt{2\sigma}} 
= \frac{E(\sqrt{3},\eta)}{q^2\sqrt{2\sigma}}  
+ f_1(\eta) \,\epsilon + f_2(\eta) \, \epsilon^2 + {\cal O}(\epsilon^3) ,
\label{eq:epsilon1}
\end{equation}
where 
\begin{widetext}
\begin{eqnarray}
f_1(\eta) & = & \frac{1}{2^{3/2}\sqrt{\pi}} \Bigg\{ 4 \sum_{j=1}^{\infty} j^2 
\left[ z_{5/2}(0,j^2\sqrt{3})  - \frac{1}{3} z_{5/2}(0,j^2/\sqrt{3}) \right] 
+ 8 \sum_{j,k=1}^{\infty} \left( k^2 - \frac{j^3}{3} \right) 
z_{5/2}(0,j^2/\sqrt{3}+k^2\sqrt{3}) \nonumber \\ & 
+& 2 \sum_{j=1}^{\infty} (-1)^j j^2 \left[ z_{5/2}((\pi\eta)^2,j^2\sqrt{3}) 
- \frac{1}{3} z_{5/2}((\pi\eta)^2,j^2/\sqrt{3}) \right] 
%\nonumber \\ & &
+ 4 \sum_{j,k=1}^{\infty} (-1)^{j+k} \left[ k^2 - \frac{j^2}{3} \right] 
z_{5/2}((\pi\eta)^2,\frac{j^2}{\sqrt{3}}+k^2\sqrt{3}) \nonumber \\ & 
+& 4 \sum_{j,k=1}^{\infty} \left[ \left( k-\frac{1}{2} \right)^2 -
\frac{1}{3} \left( j-\frac{1}{2} \right)^2 \right]
z_{5/2}(0,\eta^2+(j-1/2)^2/\sqrt{3}+(k-1/2)^2\sqrt{3}) \Bigg\} ,
\label{eq:f1big}
\end{eqnarray}
\end{widetext}
and the function $f_2$ is also explicitly known \cite{SaTr12}.
To investigate the stability of structure I, it is sufficient
to study the sign of $f_1$, which is worked out from a Taylor
expansion for small $\eta$. The first two derivatives of this 
function $f_1$ vanish at $\eta=0$ and we have 
$f_1(\eta) = - 0.5833059875\ldots\eta^2 + {\cal O}(\eta^4)$,
hence an energy gain upon increasing $\epsilon$ as compared 
to the $\epsilon=0$ case (structure I). This implies that
$\eta_{\mathrm{\tiny I}}=0$: at finite but small distances $\eta$, the
optimal phase is not structure I. To obtain the optimal value of
$\epsilon$ selected and that we denote $\epsilon^*$, 
we further Taylor expand $f_2(\eta)$ which yields 
$f_2(\eta) =  0.0408440789\ldots + {\cal O}(\eta^2)$.
As a consequence, 
\begin{equation} 
\sqrt{3} - \Delta^* \equiv \epsilon^* = 
- \frac{f_1(\eta)}{2 f_2(\eta)} = 7.14064\ldots \eta^2 + {\cal O}(\eta^4) ,
\label{eq:transitionItoII}
\end{equation}
which entails that the energy change scales like $\eta^4$. 
For the thresholds $\eta_{\mathrm{\tiny I}}$ reported in Refs. 
\cite{GoPe96,WeLJ01}, a relative accuracy of $10^{-9}$ was therefore
required to answer the finite or vanishing $\eta_{\mathrm{\tiny I}}$ question.
The accuracy of our findings is illustrated in Fig. \ref{fig:enI}. 

\begin{figure}[tb]
\begin{center}
\includegraphics[width=0.4\textwidth]{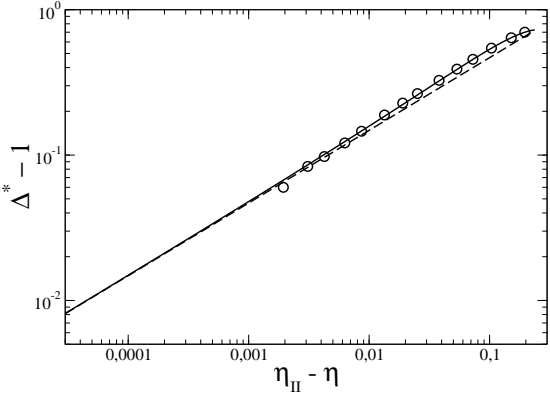}
\caption{The transition II $\to$ III: Test of the analytical 
asymptotic relation 
(\ref{eq:transitionIItoIII}) (dashed line) against numerical minimization 
of the energy (\ref{eq:series1}) (solid curve), in a log-log scale. The numerical data of Ref. \cite{GoPe96} are shown by the circles.}
\label{fig:zoom23}
\end{center}
\end{figure}

The above analysis shows that the evolution from structure I to structure 
II is not a phase transition in the common sense. The situation differs
between structures II and III. To inspect the corresponding transition,
we note that $E(\Delta,\eta)$ enjoys the symmetry $\Delta \to \Delta^{-1}$,
as is clear from Fig. \ref{fig:struct} where a global rotation of $\pi/2$
does not affect the energy but interchanges lattice vectors $\bm{a}_1$
and $\bm{a}_2$. The value $\Delta=1$ characterizing structure III
is therefore a self-dual point, and it will now be convenient
to parameterize the aspect ratio as $\Delta = \exp(\epsilon)$. 
All expressions will then be even in $\epsilon$. The
expansion of $E(e^\epsilon,\eta)$ in small $\epsilon$-deviations
yields 
\begin{equation}
\frac{E(e^{\epsilon},\eta)}{q^2\sqrt{2\sigma}} = 
\frac{E(1,\eta)}{q^2\sqrt{2\sigma}}
 + g_2(\eta) \, \epsilon^2 
+ g_4(\eta) \, \epsilon^4 + {\cal O}(\epsilon^6) ,
 \label{eq:phase23}
\end{equation}
where $g_2$ and $g_4$ are explicitly known \cite{SaTr12},
in a series form that is very reminiscent of Eq. (\ref{eq:f1big}).
The bilayer energy appears in a standard Ginzburg-Landau
form \cite{GL}, but it should be emphasized that at variance with 
mean-field arguments usually underlying such approaches, our expression
is exact. The critical point $\eta_{\mathrm{\tiny II}}$ 
sought for is the root of $g_2(\eta)=0$, which gives
$\eta_{\mathrm{\tiny II}}=0.2627602682\ldots$\
It appears here that the thresholds reported in earlier works
were accurate: $0.27$ \cite{Falk94}, 
$0.262$ \cite{GoPe96}, $0.27$ \cite{LoNe07} and $0.28$ \cite{WeLJ01}.
Proceedings along similar lines as for the I $\to$ II crossover,
we Taylor expand $g_2(\eta)$ and $g_4(\eta)$ to leading order around
$\eta_{\mathrm{\tiny II}}$. The former behaves like 
$(\eta-\eta_{\mathrm{\tiny II}})$ while the latter is constant,
a prototypical scenario for a continuous phase transition with critical
index $\beta=1/2$ \cite{GL}. Specifically, we get
\begin{equation} 
\Delta^* - 1 \simeq \epsilon^* = \left( 
- \frac{g_2(\eta)}{2 g_4(\eta)} \right)^{1/2} 
\simeq 1.48031 \sqrt{\eta_{\mathrm{\tiny II}}-\eta} .
\label{eq:transitionIItoIII}
\end{equation}
This expression applies for $\eta \leq \eta_{\mathrm{\tiny II}}$, in the
stability domain of structure II, and is in excellent agreement with our 
numerical calculations from Eq. (\ref{eq:series1}), see Fig. \ref{fig:zoom23}. 

The task remaining is to find the series representations for structures
IV and V. We first address structure IV.
Implementation of the procedure that led to the series 
(\ref{eq:series1}) becomes possible once the distance between a reference
ion and an arbitrary ion located on the same layer at 
${\bf r}(j,k) = j \bm{a}_1 + k \bm{a}_2$ is 
expressed as 
\begin{eqnarray}
&&\vert {\bf r}(j,k)\vert^2 = a^2 \left( j^2 + k^2 + 2 j k \cos\varphi \right)
\nonumber\\
&&= a^2 \left[ (j+k)^2 \cos^2(\varphi/2) + (j-k)^2 \sin^2(\varphi/2) \right] .
\end{eqnarray}
The latter ``diagonalized'' form in terms of indices, provides the
starting point to write the intra-layer Coulomb energy (summing 
$1/\sqrt{\vert {\bf r}(j,k)\vert^2}$), and suggests to introduce new
indices $n$ and $m$: if $j+k$ is even, we define $n=(j+k)/2$ and 
$m=(j-k)/2$. 
%covering all integers except for $(n,m)\ne (0,0)$.
If $j+k$ is odd, we introduce indices $n=(j+k+1)/2$ and 
$n=(j-k+1)/2$. Likewise for the inter-layer interactions, taking due account
of the shift $(\bm{a}_1+\bm{a}_2)/2$ between opposite layers.
Building on the gamma identity and Poisson summation formula,
the series form for the energy $E_{\mathrm{\tiny IV}}$ ensues \cite{SaTr12}. 
This energy depends on the angle $\varphi$ and of course on the distance 
$\eta$. 
For our purposes, rather than the lengthy explicit form,
it is sufficient to report the Landau-like expansion of $E_{\mathrm{\tiny IV}}$
in the vicinity of $\varphi=\pi/2$. A convenient expansion parameter
is $\epsilon$ such that $\exp(\epsilon) = \tan(\varphi/2)$, and the invariance
$\varphi \to \pi -\varphi$ makes $E_{\mathrm{\tiny IV}}$ 
an even function of $\epsilon$.
In the small $\epsilon$ region of interest associated to the 
vicinity of $\pi/2$ for $\varphi$, we obtain an expansion up to order 
$\epsilon^4$ of the same form as (\ref{eq:phase23}). 
This teaches us that structure III is unstable for 
$\eta > \eta_{\mathrm{\tiny III}}=0.6214809246\ldots$,
to be compared to the thresholds $0.61$ \cite{EsKa95},
$0.622$ \cite{GoPe96}, $0.62$ \cite{LoNe07}, $0.59$ \cite{WeLJ01}
while structure IV was not considered in \cite{Falk94}.
We furthermore again obtain a second order phase transition with critical
index $1/2$ and explicit order parameter close to the transition point
\begin{equation} 
\epsilon^* \simeq \frac{\pi}{2} - \varphi^* 
\simeq 1.24494 \sqrt{\eta-\eta_{\mathrm{\tiny III}}},
\label{eq:transitionIIItoIV}
\end{equation}
in excellent agreement with our numerical data.

The transition  IV $\to$ V is discontinuous, which made its
characterization more elusive in previous publications. 
Our method, though, is easily adapted to the geometry of structure V.
%once it has been remarked that
%\begin{equation}
%\frac{\vert {\bf r}(j,k)\vert^2}{a^2} = j^2 + k^2 + j k 
%= \frac{ 3 (j+k)^2 + (j-k)^2 }{4} 
%\end{equation}
%for the ``intra'' distance between points $(0,0)$ and 
%$(j,k)$ whereas when the two points belong to opposite surfaces,
%a little algebra leads to the ``inter'' form
%($d$ being the plate separation)
%\begin{equation}
%\vert {\bf r}(j,k)\vert^2 = 
%a^2 \left[ (j+1/3)^2 + (k+1/3)^2 + (j+1/3) (k+1/3) \right] + d^2 =
%\frac{a^2}{4} \left[ 3 (j+k+2/3)^2 + (j-k)^2 \right] + d^2 .
%\end{equation}
The series representation for $E_{\mathrm{\tiny V}}(\eta)$ should be compared 
to $E_{\mathrm{\tiny IV}}(\varphi^*,\eta)$ evaluated for the
optimal distortion angle $\varphi^*(\eta)$. 
Requiring that $E_{\mathrm{\tiny V}}(\eta)=E_{\mathrm{\tiny IV}}(\varphi^*,\eta)$,
we obtain the last $\eta$-threshold that was remaining to be specified:
$\eta_{\mathrm{\tiny IV}} = 0.73242\ldots$\ Previous investigations
gave $0.75$ \cite{EsKa95}, $0.732$ \cite{GoPe96}, $0.87$ \cite{LoNe07} and
$0.70$ \cite{WeLJ01}. For $\eta>\eta_{\mathrm{\tiny IV}}$, structure V
is energetically favorable. As a by-product of our analysis,
we report the large distance behavior of the interplate pressure
$P = -2\sigma \, \partial E_{\mathrm{\tiny V}}/\partial d$:
\begin{equation}
P \, = \,  -6 \pi (\sigma q)^2 \, 
%\frac{3\sqrt 2}{4} 
\exp
\left(-\frac{4 \pi}{\sqrt 2 \, 3^{1/4}} \, \eta\right), 
\end{equation}
in agreement with \cite{GoPe96} but at variance with \cite{EsKa95,LaLP00}.
%XXXX errors in litterature $3 \sqrt 2/4 \simeq 1.06$ \cite{LaLP00}

\begin{figure}[htb]
\includegraphics[width=0.4\textwidth]{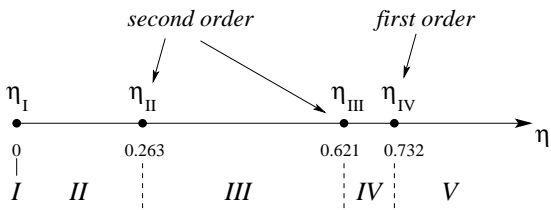} 
\caption{{Sequence of structures encountered as a function of
reduced inter-plate separation $\eta$. The values reported for the
different stability thresholds are rounded to the third digit.}}
\label{fig:summary}
\end{figure}

To summarize, we have derived series representations for the different Coulomb 
lattice sums pertaining to the ground state of classical bilayer systems. 
The derivation, worked out explicitly for the five different structures 
that were known to compete at vanishing temperature, results from a series 
of transformation rooted in the general theory of Jacobi $\theta$ functions 
\cite{SaTr12}. 
The resulting series provide the thresholds delimiting the domains of 
validity of the different phases, that were prone to some fluctuations 
in previous works. {Figure \ref{fig:summary} provides an overview of our 
main findings.}
We could in particular show that the simple hexagonal structure I can only 
exist in the limiting case of a vanishing interplate distance, 
and is preempted by a buckled phase for all $\eta \neq 0$. 
This is the scenario first reported in Ref. \cite{Messina03},
which differs from several other studies that assigned a finite stability
window to phase I.
%We have thus clarified incompatible results published in the literature. 
Whereas the evolution I $\to$ II is not a phase transition, we could show 
that the continuous transitions II $\to$ III and III $\to$ IV have critical 
index $\beta=1/2$.
In addition, our series representation is endowed with exceptional convergence 
properties, providing typically more than 10 digits of accuracy when retaining
only the first 4 or 5 terms involved. 
Relinquishing the symmetry between the two plates to address the cases where 
they bear different surface charges is an interesting venue for future work. 
This brings the difficulty that local electroneutrality no longer holds at 
the single-plate level in the ground state \cite{Messina00}, 
except presumably at large separations. Our approach can also be 
extended to bilayers and multilayers with repulsive Yukawa or inverse-power-law interactions,
that deserve attention. 

% summary of all threshold values ?

\begin{acknowledgments}
We would like to thank M. Mazars and C. Texier for useful discussions.
The support received from Grants VEGA No. 2/0049/12 and CE-SAS QUTE
is acknowledged. 
\end{acknowledgments}


\begin{thebibliography}{1}

\bibitem{Wi34}
E. Wigner, Phys. Rev. {\bf 46}, 1002 (1934). 

\bibitem{GrAd79}
{C.G. Grimes and G. Adams, 
Phys. Rev. Lett. {\bf 42}, 795 (1979).}

%GaAs
\bibitem{Shaye}
H. C. Manoharan, Y. W. Suen, M. B. Santos, and M. Shayegan, 
Phys. Rev. Lett. {\bf 77}, 1813 (1996);
E. Tutuc, M. Shayegan, and D. A. Huse,
Phys. Rev. Lett. {\bf 93}, 036802 (2004).

%excitons in semiconductors
\bibitem{Eise04}
J. P. Eisenstein and A. H. MacDonald,
Nature {\bf 432}, 691 (2004).

%quantum dots
\bibitem{ImMA96}
H. Imamura, P. A. Maksym and H. Aoki, 
Phys. Rev. B {\bf 53}, 12 613 (1996).

%bilayer graphene
\bibitem{graphene}
A. S. Mayorov, D. C. Elias, M. Mucha-Kruczynski, R. V. Gorbachev, 
T. Tudorovskiy, A. Zhukov, S. V. Morozov, M. I. Katsnelson, V. I. Falko,
A. K. Geim and K. S. Novoselov, Science {\bf 6044}, 860 (2011).

%Boron Nitride bilayers
\bibitem{boron}
R. M. Ribeiro and N. M. R. Peres,
Phys. Rev. B {\bf 83}, 235312 (2011).

%plasmas
\bibitem{Mitc98}
T. B. Mitchell, J. J. Bollinger, D. H. E. Dubin, X.-P. Huang, W. M. Itano, 
B. M. Baughman, Science {\bf 282}, 1290 (1998).

%dusty plasmas
\bibitem{dusty}
L. Teng and P. Tu, L. I, 
Phys. Rev. Lett. {\bf 90}, 245004 (2003).

%colloids
\bibitem{colloids}
S. Neser, C. Bechinger, P. Leiderer and T. Palberg,
Phys. Rev. Lett. {\bf 79}, 2348 (1997).	

\bibitem{rque0}
Note though that in the soft matter realm, the charge carriers need to
be multivalent in order to reach, at room temperature and in 
an aqueous dispersion, the strong couplings required 
for Wigner crystallization \cite{Grosberg02,Naji}.

\bibitem{Falk94}
V.I. Falko,
Phys. Rev. B {\bf 49}, 7774 (1994).

\bibitem{EsKa95}
K. Esfarjani and Y. Kawazoe,
J. Phys.: Condens. Matter {\bf 7} 7217 (1995).

\bibitem{GoPe96}
{G. Goldoni and F. M. Peeters,
Phys. Rev. B {\bf 53}, 4591 (1996).}

\bibitem{Messina03}
{R. Messina and H. L\"owen, 
Phys. Rev. Lett. {\bf 91}, 146101 (2003); E. C. O\v{g}uz,
R. Messina, and H. L\"owen, Europhys. Lett. {\bf 86}, 28002 (2009).}

\bibitem{LoNe07}
{V. Lobaskin and R. R. Netz,
Europhys. Lett. {\bf 77}, 38003 (2007).}

\bibitem{ScSP99}
{I. V. Schweigert, V. A. Schweigert, and F. M. Peeters, 
Phys. Rev. Lett. {\bf 82}, 5293 (1999); Phys. Rev. B {\bf 60}, 14 665
(1999).}

\bibitem{WeLJ01}
{J. J. Weis, D. Levesque, and S. Jorge,
Phys. Rev. B {\bf 63}, 045308 (2001).}

\bibitem{Maza11}
{M. Mazars, Phys. Rep. {\bf 500}, 43 (2011).}

\bibitem{Levin02}
{Y. Levin, Rep. Prog. Phys. {\bf 65}, 1577 (2002).}

\bibitem{Naji}
A. Naji, M. Kanduc, R. R. Netz and Rudolf Podgornik,
in Understanding Soft Condensed Matter via Modeling and Computation, 
Eds. W.-B. Hu and A.-C. Shi (World Scientific, Singapore, 2010).

\bibitem{LaLP00}
{A. W. C. Lau, D. Levine, and P. Pincus,
Phys. Rev. Lett. {\bf 84}, 4116 (2000).}

\bibitem{SaTr11}
{L. \v{S}amaj and E. Trizac, 
Phys. Rev. Lett. {\bf 106}, 078301 (2011);
Phys. Rev. E {\bf 84}, 041401 (2011).}

\bibitem{Grosberg02}
{A. Y. Grosberg, T. T. Nguyen, and B. I. Shklovskii,
Rev. Mod. Phys. {\bf 74}, 329 (2002).}

\bibitem{TrVa06}
A. Travesset and D. Vaknin,
Europhys. Lett. {\bf 74}, 181 (2006).

\bibitem{rque}
Indeed, Earnshaw theorem 
[S. Earnshaw, Trans. Camb. Phil. Soc., {\bf 7}, 97 (1842)]
states that a collection of point charges cannot be maintained 
in stable equilibrium under the action of electrostatic forces alone. Hence,
at $T=0$, the charges stick to the boundaries, and they do so evenly. 
The particle surface density on each plate is then $\sigma$,
so that electroneutrality holds at the level of each plate. 
For asymmetric plates, the density of particles may 
violate neutrality at the plate level, global neutrality
being always enforced \cite{Messina00}. {A similar argument can
be found in footnote 27 of Ref. \cite{Messina09}}.

\bibitem{rque101}
{Clearly, the ground state structure does not depend on the precise 
value of the confining plates surface charge, as long as they
are symmetric objects, thereby producing a vanishing electric
field in the interstitial slab \cite{Messina09}.}

\bibitem{Messina09}
{R. Messina, J. Phys.: Condens. Matter {\bf 21}  113102 (2009).}

% 5 structures in quantum context
\bibitem{NaHo95}
S. Narasimhan and T.-L. Ho 
Phys. Rev. B {\bf 52}, 12291 (1995).

%Wigner monolayer -> hexagonal
\bibitem{BoMa77}
L. Bonsall and A. A. Maradudin,
Phys. Rev. B {\bf 15}, 1959 (1977).

%Ewald
\bibitem{dLPS80} S. W. de Leeuw, J. W. Perram and E. R. Smith,
Proc. R. Soc. Lond. A, {\bf 373}, 27 (1980).

\bibitem{SaTr12}
L. \v{S}amaj and E. Trizac,
in preparation.

\bibitem{Misra40}
R. D. Misra,
Proc. Camb. Phil. Soc. {\bf 36}, 173 (1940);
M. Born and R.D. Misra,
Proc. Camb. Phil. Soc. {\bf 36}, 466 (1940).

\bibitem{Borwein88}
{D. Borwein, J. M. Borwein, R. Shail, and I. J. Zucker,
J. Phys. A: Math. Gen. {\bf 21}, 1519 (1988).}

\bibitem{Bowick06}
{M. J. Bowick, A. Cacciuto, D. R. Nelson, and A. Travesset,
Phys. Rev. B {\bf 73}, 024115 (2006).}

\bibitem{GL}
L. D. Landau and E. M. Lifshitz,
Statistical Physics, Course of Theoretical Physics vol 5,
Pergamon Press (1980).

\bibitem{Messina00}
{R. Messina, C. Holm, and K. Kremer, 
Phys. Rev. Lett. {\bf 85}, 872 (2000).
R. Messina, C. Holm and K. Kremer,  Europhys. Lett.
{\bf 51}, 461 (2000).}

\end{thebibliography}
\end{document}